\documentclass[a4paper,11pt]{article}

\usepackage{lineno}

\usepackage{pos}

\title{Physics Performance of the Large-Sized Telescope prototype of the Cherenkov Telescope Array}
 \ShortTitle{Physics Performance LST}

\author*[a,b]{R. L\'opez-Coto}
\author[c]{A. Moralejo}
\author[c]{M. Artero}
\author[d]{A. Baquero}
\author[a,e]{M. Bernardos}
\author[d]{J.L. Contreras}
\author[f]{F. Di Pierro}
\author[g]{E. Garc\'ia}
\author[c]{D. Kerszberg}
\author[d]{M. L\'opez-Moya}
\author[d]{A. Mas Aguilar}
\author[d]{D. Morcuende}
\author[h]{M. Noethe}
\author[i]{S. Nozaki}
\author[j]{Y. Ohtani}
\author[c]{C. Priyadarshi}
\author[k]{Y. Suda}
\author[g]{T. Vuillaume}
\author[]{for the CTA LST project}

\affiliation[a]{Istituto Nazionale di Fisica Nucleare, Sezione di Padova, I-35131, Padova, Italy.}
\affiliation[b]{now at Instituto de Astrof\'isica de Andaluc\'ia, CSIC, 18080 Granada, Spain.}
\affiliation[c]{Institut de Fisica d'Altes Energies (IFAE), The Barcelona Institute of Science and Technology, Campus UAB, 08193 Bellaterra (Barcelona), Spain.}
\affiliation[d]{IPARCOS and Department of EMFTEL, Universidad Complutense de Madrid, E-28040 Madrid}
\affiliation[e]{Universit\`a degli Studi di Padova, Via Marzolo 8, 35131, Padova, Italy.}
\affiliation[f]{INFN Sezione di Torino, via P. Giuria, 1 - 10125 Torino, Italy.}
\affiliation[g]{LAPP, Univ. Grenoble Alpes, Univ. Savoie Mont Blanc, CNRS-IN2P3, 74000 Annecy, France, 9 Chemin de Bellevue - BP 110, 74941 Annecy Cedex, France.}
\affiliation[h]{Department of Physics, TU Dortmund University, Otto-Hahn-Str. 4, 44221 Dortmund, Germany.}
\affiliation[i]{Division of Physics and Astronomy, Graduate School of Science, Kyoto University, Sakyoku, Kyoto, 606-8502, Japan.}
\affiliation[j]{Institute for Cosmic Ray Research, University of Tokyo, 5-1-5, Kashiwa-no-ha, Kashiwa, Chiba 277-8582, Japan.}
\affiliation[k]{Max-Planck-Institut f{\"u}r Physik, 80805 M{\"u}nchen, Germany.}

\emailAdd{rlopez@pd.infn.it}

\abstract{The Large-Sized Telescope (LST) prototype of the future Cherenkov Telescope Array (CTA) is located at the Northern site of CTA, on the Canary Island of La Palma. It is designed to provide optimal performance in the lowest part of the energy range covered by CTA, observing gamma rays down to energies of tens of GeV. The LST prototype started performing astronomical observations in November 2019 during the commissioning of the telescope and it has been taking data since then. In this contribution, we will present the tuning of the characteristics of the telescope in the Monte Carlo (MC) simulations to describe the data obtained, the estimation of its angular and energy resolution, and an evaluation of its sensitivity, both with simulations and with observations of the Crab Nebula.}

\FullConference{37$^{\rm{th}}$ International Cosmic Ray Conference (ICRC 2021)\\
		July 12th -- 23rd, 2021\\
		Online -- Berlin, Germany}

\begin{document}
\maketitle

\section{Introduction}
\label{sec:intro}
The Cherenkov Telescope Array (CTA) will consist of two arrays of imaging atmospheric Cherenkov telescopes (IACTs) located at the Northern and Southern Hemispheres \citep{general_cta}. The arrays will be composed of IACTs of different sizes, optimised for different energy ranges within the Very-High Energy (VHE) band. The largest telescopes of the array will be the Large-Sized Telescopes (LSTs), with a focal length of 28 m and a parabolic mirror dish of 23 m diameter. LSTs target the lowest energies accessible to CTA, down to a threshold energy of $\simeq$20 GeV, and have a light carbon-fiber structure for fast repositioning to enable the follow-up of short transient events. The camera comprises 1855 photomultiplier tubes that transform photons into electrical signals, recorded by a fast readout system digitizing at 1 Gsample/s \citep{general_cta}.  

LST-1, the first LST \cite{general_lst}, is located at the CTA-north site, at a height of 2200 m a.s.l. at the Roque de los Muchachos Observatory on the Canary Island of La Palma, Spain (28$^\circ$N, 18$^\circ$W). The first LST-1 sky observations took place in December 2018, and the telescope has been taking commissioning data since then. 
We evaluate the performance of the LST-1 using Monte Carlo simulations and observations of the Crab Nebula, the standard candle of VHE $\gamma$-ray astronomy. Located in our Galaxy at $\sim$ 2 kpc from the Earth, the Crab is the best-studied pulsar wind nebula (PWN), and thus considered the archetypal plerion \citep{CrabMAGIC}.

\section{Data analysis}
\label{sec:analysis}
We perform the analysis of the LST-1 data using \texttt{cta-lstchain}\footnote{https://github.com/cta-observatory/cta-lstchain}, a custom pipeline heavily based on \texttt{ctapipe}\footnote{https://github.com/cta-observatory/ctapipe}, the framework for prototyping the low-level data processing algorithms for CTA. The main input of \texttt{cta-lstchain} are the camera events written to disk by the data acquisition (DAQ), containing pixel-wise waveforms, which are calibrated and integrated to obtain the charge (in photo-electrons) and arrival time \cite{calibration} of the Cherenkov signal in each pixel. The image cleaning procedure selects pixels with signals dominated by Cherenkov light from the recorded shower, rather than background light of the night sky. The resulting clean image is then characterized by the extended Hillas parameters, including the moments up to 3$^{rd }$order of the light distribution of the images. The calibrated and parametrized images constitute the Data Level 1 (DL1). 

We use Random Forest models, fed with image parameters and trained on Monte Carlo simulations \cite{cta_mc} to select $\gamma$-ray showers among the overwhelming background of hadron-initiated showers, and to estimate the arrival direction and the energy of the primary gamma rays. These models are then applied to the recorded events; the output of this stage, the DL2 data level, consists of event lists containing the reconstructed shower parameters: energy, direction and \emph{gammaness}, a score in the range 0 to 1, ranging from least- to most-gamma-like, which allows us to select samples of gamma-like events with different levels of residual background.

Within this basic scheme, we perform two different types of analysis:

\paragraph{Source-independent analysis}
The source-independent analysis only uses image parameters that can be calculated without any a priori knowledge of the source position in the camera. This is the standard analysis that is used in stereoscopic IACT observations.

\paragraph{Source-dependent analysis}

Assuming prior knowledge of the source position in the camera has advantages, especially for the case of single-telescope observations, in which it is difficult to reconstruct the shower geometry unambiguously. For the case of a single point-like source in the field of view, for which all gamma rays are expected to arrive from the same direction, source-dependent analysis provides better reconstruction of the shower impact parameter, hence of the gamma-ray energy and of the identity of the primary.
This analysis also has disadvantages, such as the impossibility of using it for the analysis of significantly extended sources, but it is a convenient approach for targets such as active galactic nuclei (AGNs) or the Crab Pulsar and Crab Nebula, for which only a moderate extension was measured by H.E.S.S.

\paragraph{DL3 and IRF calculation}
Finally, a cut in \emph{gammaness} is applied to select gamma-ray candidates, and instrument response functions (IRFs) are computed for the same cuts using MC simulations. The list of candidate gammas and the IRFs are the components of the Data Level 3 (DL3) which is the input for high-level analysis tools such as \texttt{gammapy}\footnote{https://github.com/gammapy/gammapy}. The gamma-ray event lists from the observed data are produced using a \texttt{cta-lstchain} specific tool while the IRFs are produced using the \texttt{pyirf}\footnote{https://github.com/cta-observatory/pyirf} library, a prototype for IRF generation for CTA. These files are generated in accordance with \texttt{gamma-astro-data-format}\footnote{https://gamma-astro-data-formats.readthedocs.io/en/latest} which is an open and interoperable data format for gamma-ray astronomy. The DL3 files are currently produced with a fixed selection cut on \emph{gammaness} for events of all energies.

The performance parameters presented in this contribution are based on the baseline analysis described above. The application of more sophisticated methods to the reconstruction of LST data, like those based on convolutional neural networks, is presented elsewhere in these proceedings (see e.g. \cite{gammalearn}).

\section{Monte Carlo (MC) simulations}
Simulations play a substantial role in the analysis of IACT data, allowing us to infer the properties of the primary gamma ray from the observed shower images. It is therefore important to validate the suitability of the simulations for the actual conditions of the observations through detailed data vs. MC comparisons.

\subsection{Absolute optical efficiency using muon rings}
\begin{figure}
\begin{center}
\includegraphics[width=\textwidth]{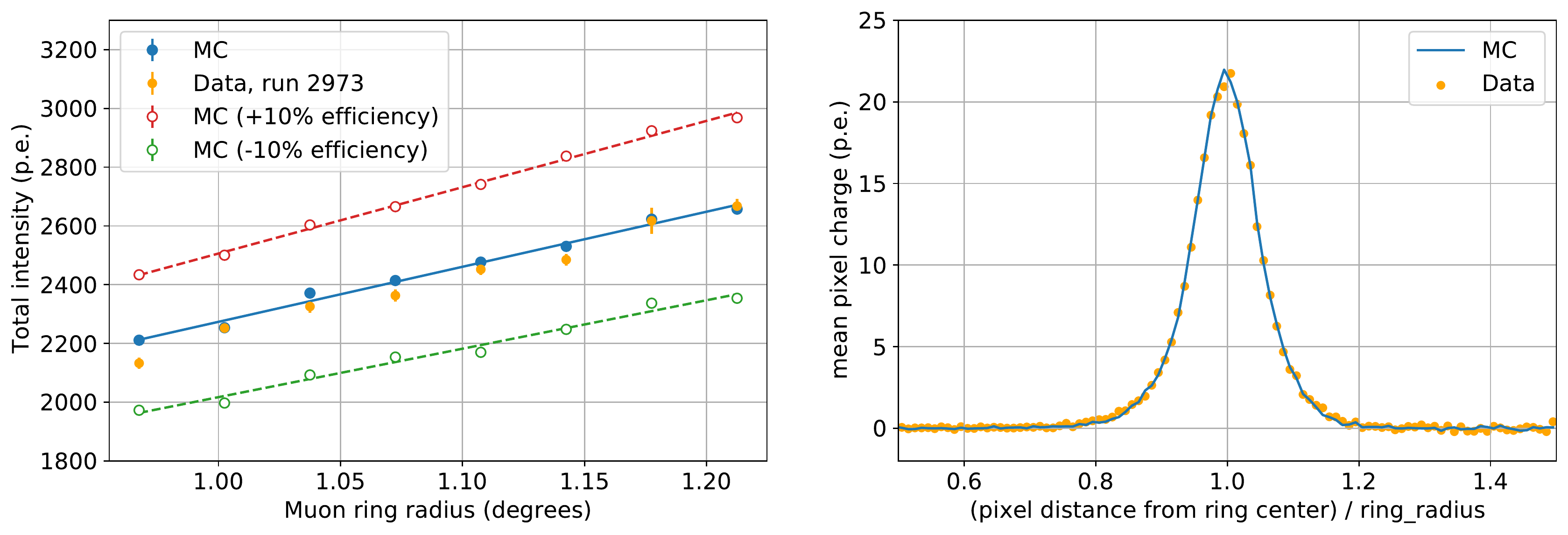}
\vspace{-0.3cm}
\caption{Left: total light in muon rings detected by LST-1 vs. ring radius. Data from one run is compared to three different simulations, with telescope efficiency changing in steps of 10$\%$. Right: average radial charge profile of muon rings, in units of the ring radius.}
\label{fig:muon}
\end{center}
\end{figure}

IACTs use camera calibration systems to estimate the conversion between the measured signal and the total number of photons detected in a pixel \cite{muons_cta, calibration}. To perform an absolute calibration of the overall light throughput of the telescope we use a signal of predictable intensity, namely the ring-shaped images produced by cosmic-ray muons. The comparison of the observed rings with those in MC simulations produced with different assumed telescope optical efficiencies allows this key parameter of an IACT to be tuned (Fig. \ref{fig:muon}, left). Furthermore, the analysis of the radial average intensity profile of muon rings provides validation (and allows fine tuning) of the simulated point-spread function of the LST-1 optics (see right panel of Fig. \ref{fig:muon}).

\begin{figure}
\begin{center}
\includegraphics[width=\columnwidth]{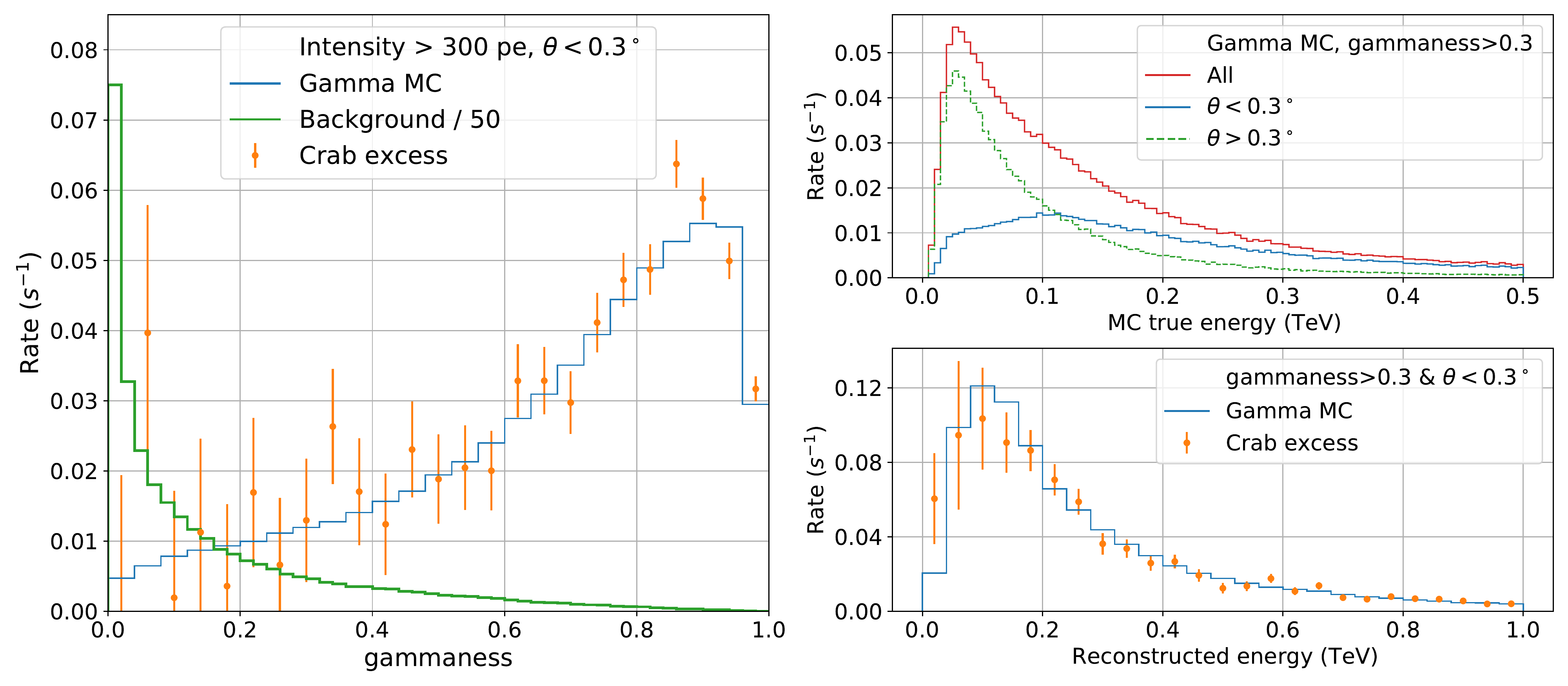}
\vspace{-0.3cm}
\caption{Left: distribution of the signal selection parameter \emph{gammaness} for images with total intensity above 300 pe. Top right: distribution of the true energy of MC gammas after a soft background rejection cut. With a single telescope, most gammas below 0.1 TeV are reconstructed more than 0.3$^\circ$ away from the source direction. Bottom right: distribution of reconstructed energy for simulated and observed gammas from the Crab Nebula. Simulated gammas are weighted to the Crab spectrum in \cite{CrabMAGIC}}.
\label{fig:data_vs_mc}
\end{center}
\vspace{-0.7cm}
\end{figure}

\subsection{Crab data / gamma MC comparisons \label{datamccomp}}
\begin{figure}
\begin{center}
\includegraphics[width=\textwidth]{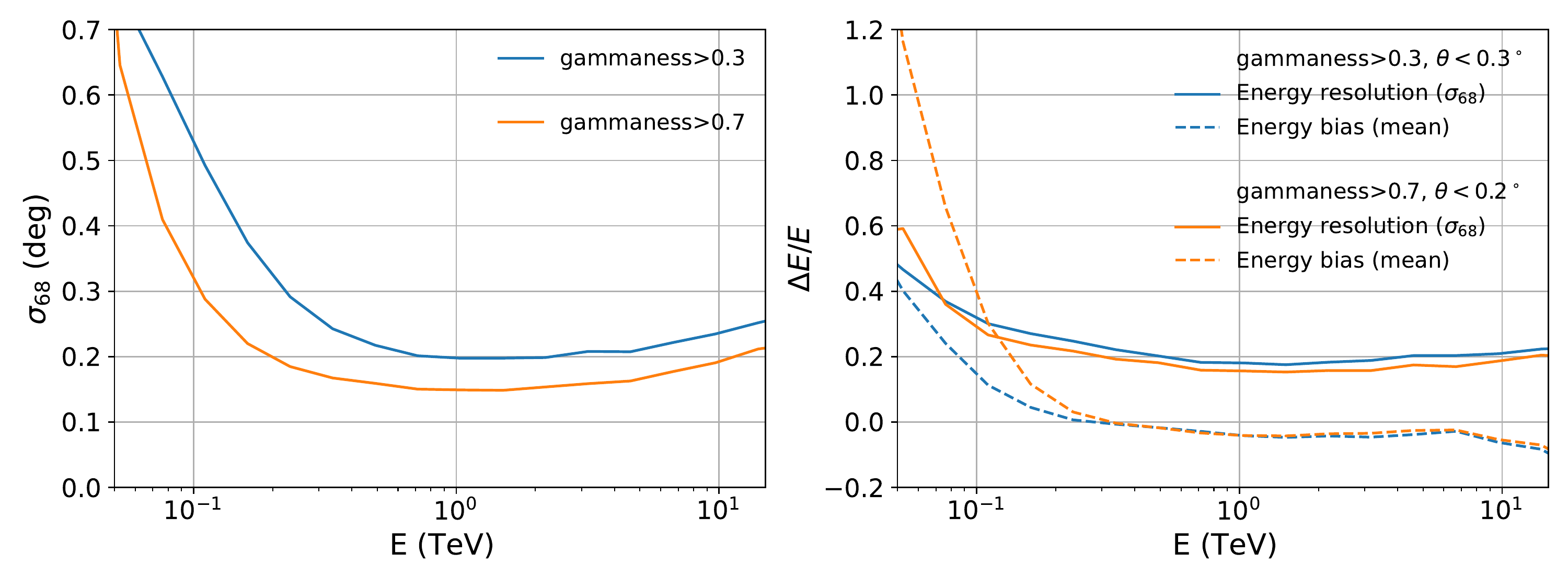}
\vspace{-0.3cm}
\caption{Angular and spectral resolutions of LST-1 vs. true energy (source-independent analysis) obtained with simulations tuned to Crab observations. Two different values of the gamma-ray selection cut are shown.}
\label{fig:resolutions}
\end{center}
\vspace{-0.7cm}
\end{figure}

The agreement between the simulations and the actual telescope performance can also be tested using the Crab Nebula as a standard candle: its high flux allows a fast significant detection of a gamma-ray excess above the fluctuations of the cosmic-ray background without using any gamma-ray selection method, just comparing the detection rates in a region around the source and those in a control region of similar acceptance. Assuming a certain energy spectrum (taken here from \cite{CrabMAGIC}), we can obtain from the MC simulations the expected distribution of any image parameter for gamma rays, and compare it to the one of the observed Crab excesses. The left panel of Fig. \ref{fig:data_vs_mc} shows, for the source-independent analysis, a comparison of the \emph{gammaness} parameter distribution for images brighter than 300 p.e. (which corresponds to a peak gamma-ray energy of 110 GeV), for a 3.5-h-long observation of the Crab Nebula in November 2020 at a zenith angle $<30^\circ$. Note that the \emph{gammaness} parameter results from a combination of all the basic image parameters (image intensity, width, length, time profile of the image, etc.), so a good agreement of its distribution in data and MC points towards similar distributions and correlations of the various image parameters. The distribution of gamma-ray rates vs. reconstructed energy (bottom right panel of Fig. \ref{fig:data_vs_mc}) also shows a good match. The energy and angular resolution derived from the same MC simulations are shown in Fig. \ref{fig:resolutions}. 

Regarding the energy threshold, the peak of the true energy of reconstructed gamma rays (for the simulated Crab spectrum) is around 30 GeV, as seen in the top right panel of Fig. \ref{fig:data_vs_mc}. In the same figure we see how, with a single telescope, the direction reconstruction for sub-100 GeV gamma images, has limited precision and most of them are further than 0.3$^\circ$ from the source.

\section{Crab Nebula spectrum and LST-1 flux sensitivity}
\begin{figure}
\begin{center}
\includegraphics[width=\textwidth]{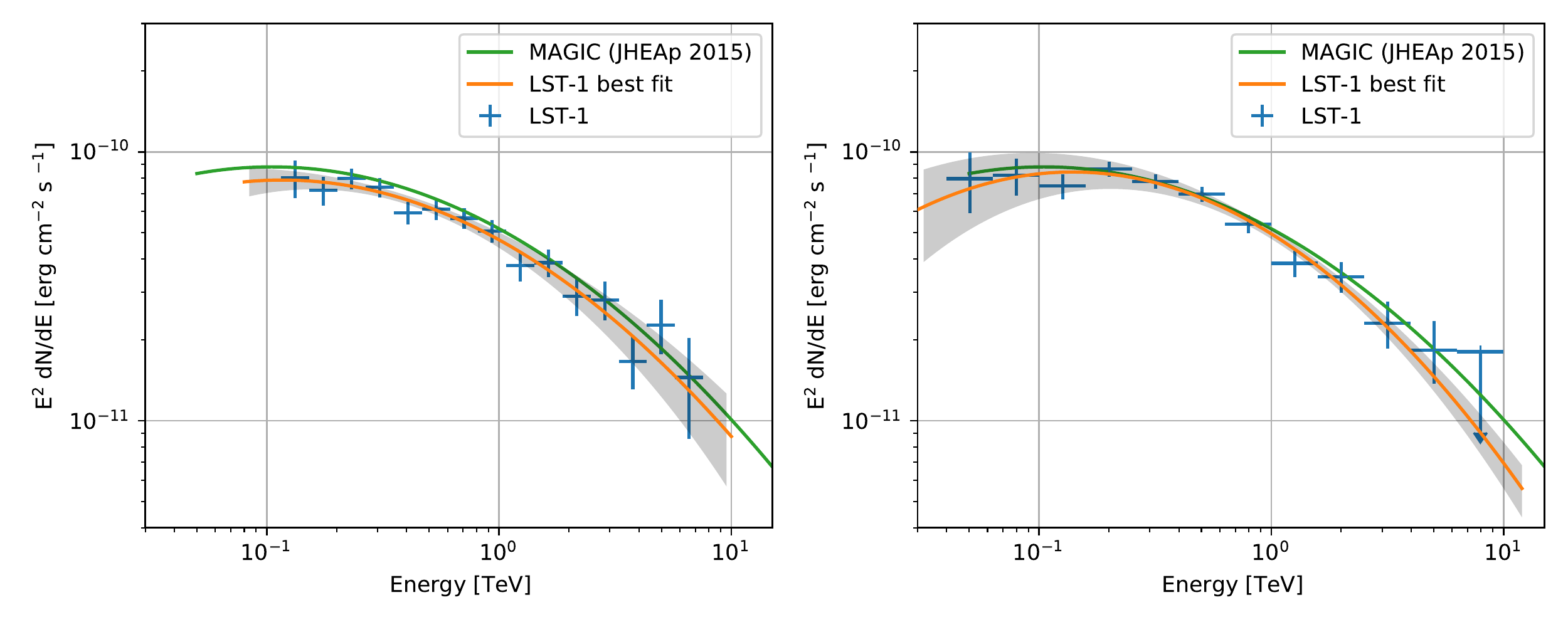}
\vspace{-0.7cm}
\caption{Spectral energy distribution of the Crab Nebula obtained with LST-1 in a 3.5-h-long observation in November 2020. Left panel: results using source-independent analysis. Right panel: results using source-dependent analysis.}
\label{fig:SEDS}
\end{center}
\vspace{-0.3cm}
\end{figure}

We used \texttt{gammapy} to process the DL3 files obtained from the processing of 3.5 hours of low-zenith observations of the Crab Nebula taken on the night of the 20$^{\rm th}$ of November 2020, a so-called \emph{reference night}, selected because of the good weather conditions and good overall performance of the telescope. The results for the source-dependent and the source-independent analyses are shown in Fig. \ref{fig:SEDS}. The spectrum derived using the source-dependent analysis extends to lower energies than the one using the source-independent analysis, since the former avoids the issues with direction reconstruction below 100 GeV pointed out in Section \ref{datamccomp}. A reference Crab spectrum published by the MAGIC collaboration is shown for comparison.

\subsection{Flux sensitivity}
\begin{figure}
\begin{center}
\includegraphics[width=0.50\columnwidth]{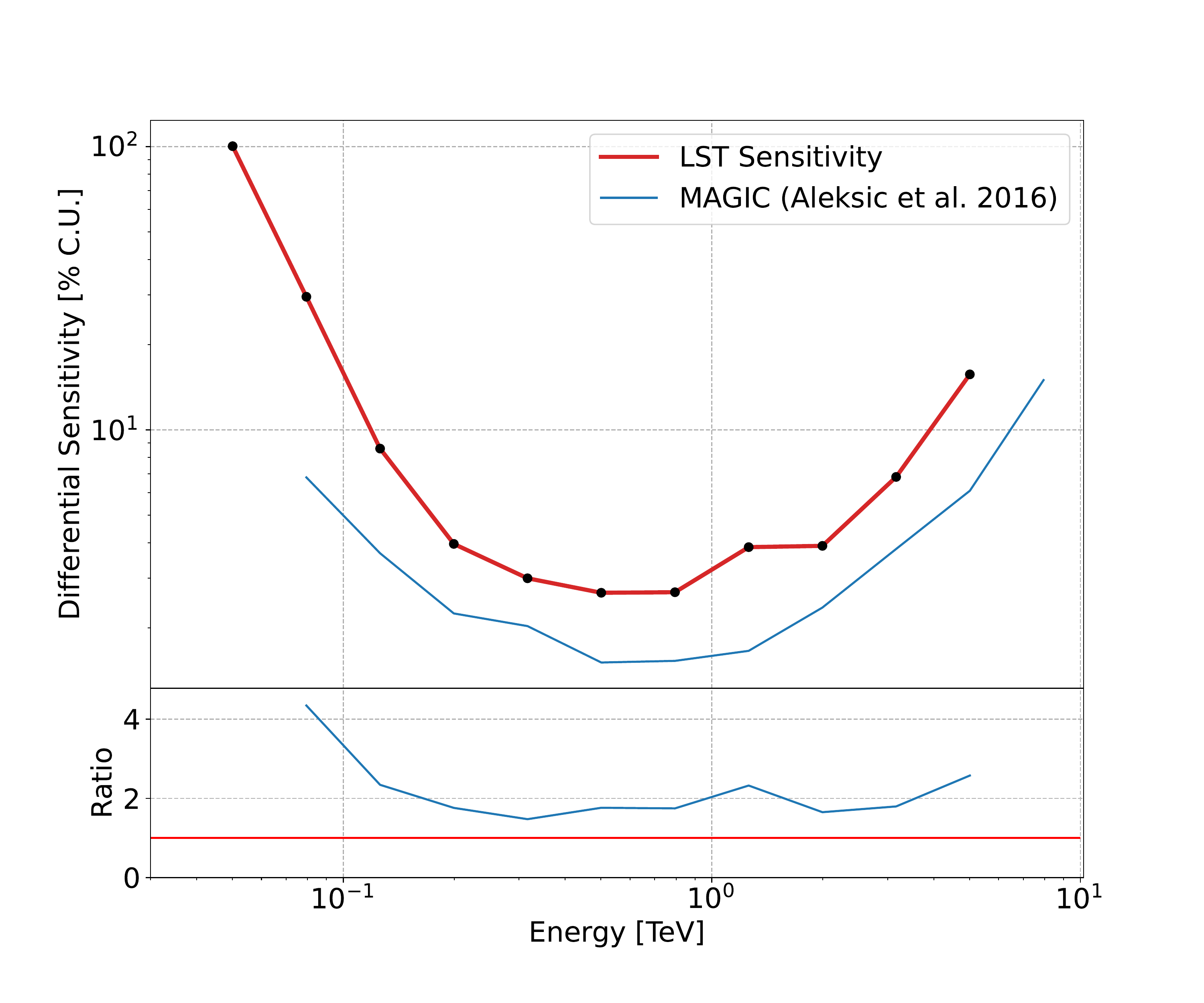}
\includegraphics[width=0.48\columnwidth]{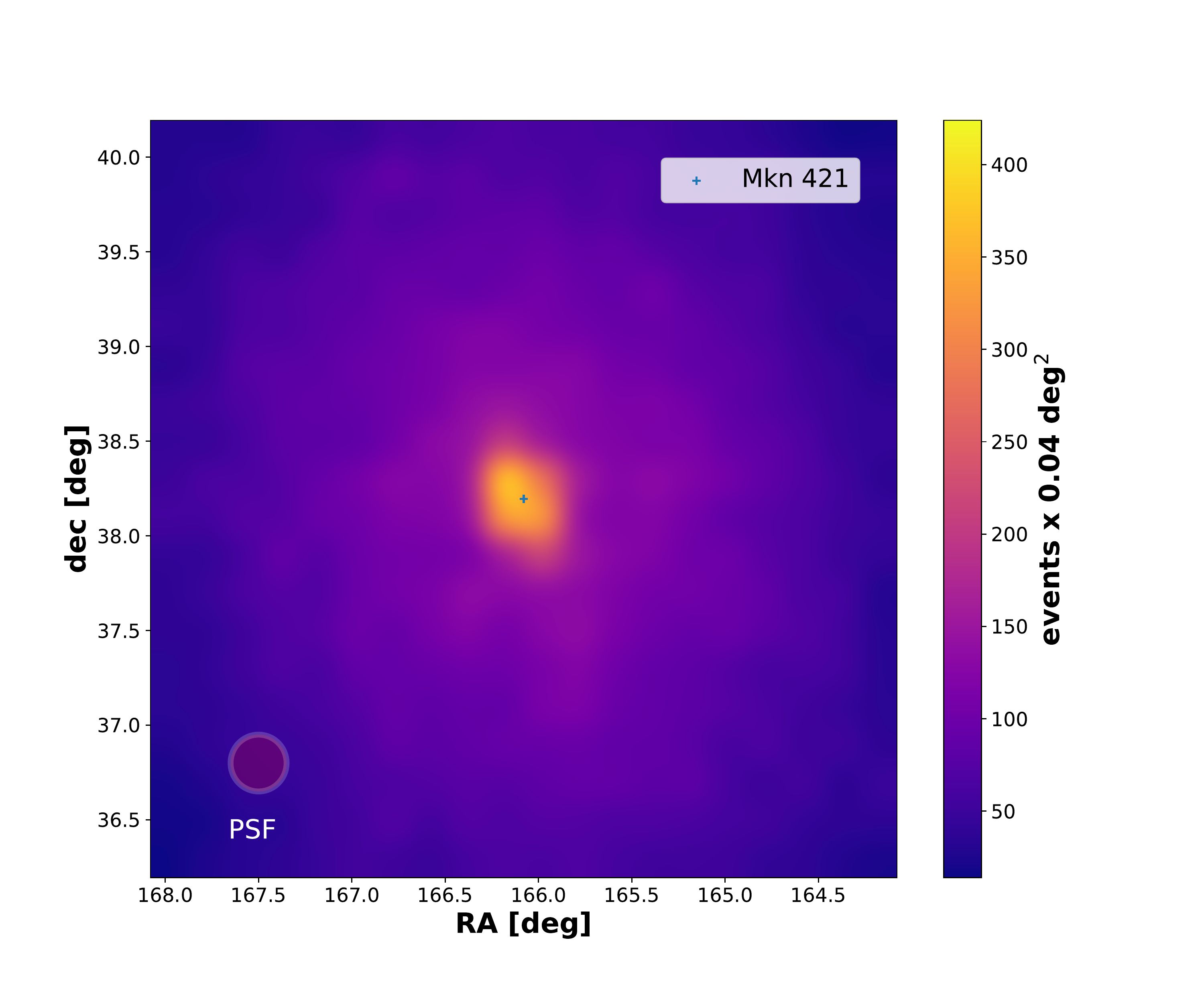}
\vspace{-0.3cm}
\caption{Left panel: LST-1 Single-telescope differential sensitivity using source-dependent analysis compared to that of MAGIC. The lower panel shows the ratio between LST single-telescope and MAGIC stereo sensitivity. Right panel: Smoothed skymap of gamma-like events recorded around Markarian 421.}
\label{fig:sensitivity}
\end{center}
\vspace{-0.7cm}
\end{figure}

The sensitivity of an IACT is defined as the minimum flux from  point-like targets that it is able to detect with a 5-$\sigma$ significance in a given time, with additional constraints on the number of excess events and signal-to-noise ratio \footnote{as defined in https://www.cta-observatory.org/science/cta-performance/}. We calculated the sensitivity by optimizing the gamma selection cuts using a large dataset of wobble observations, and applying these cuts to the data from the reference night. The result, for the source-dependent analysis, is shown in Fig. \ref{fig:sensitivity}. The MAGIC sensitivity is shown for comparison: despite the larger mirror area of the LST-1 compared to one of the two MAGIC telescopes, the advantages of stereoscopic reconstruction can be clearly seen in the graph as a factor $\sim$2 better sensitivity for MAGIC. 

\section{Other sources detected by LST-1 in its commissioning phase}
As mentioned in Section \ref{sec:intro}, the goal of the LST is to reach the lowest possible energy threshold. The observation of sources like pulsars or distant extragalactic objects benefits greatly from this low threshold. In the following, we show a few of these examples.

\subsection{Active Galactic Nuclei (AGNs)}
\begin{figure}
\begin{center}
\includegraphics[width=0.49\columnwidth]{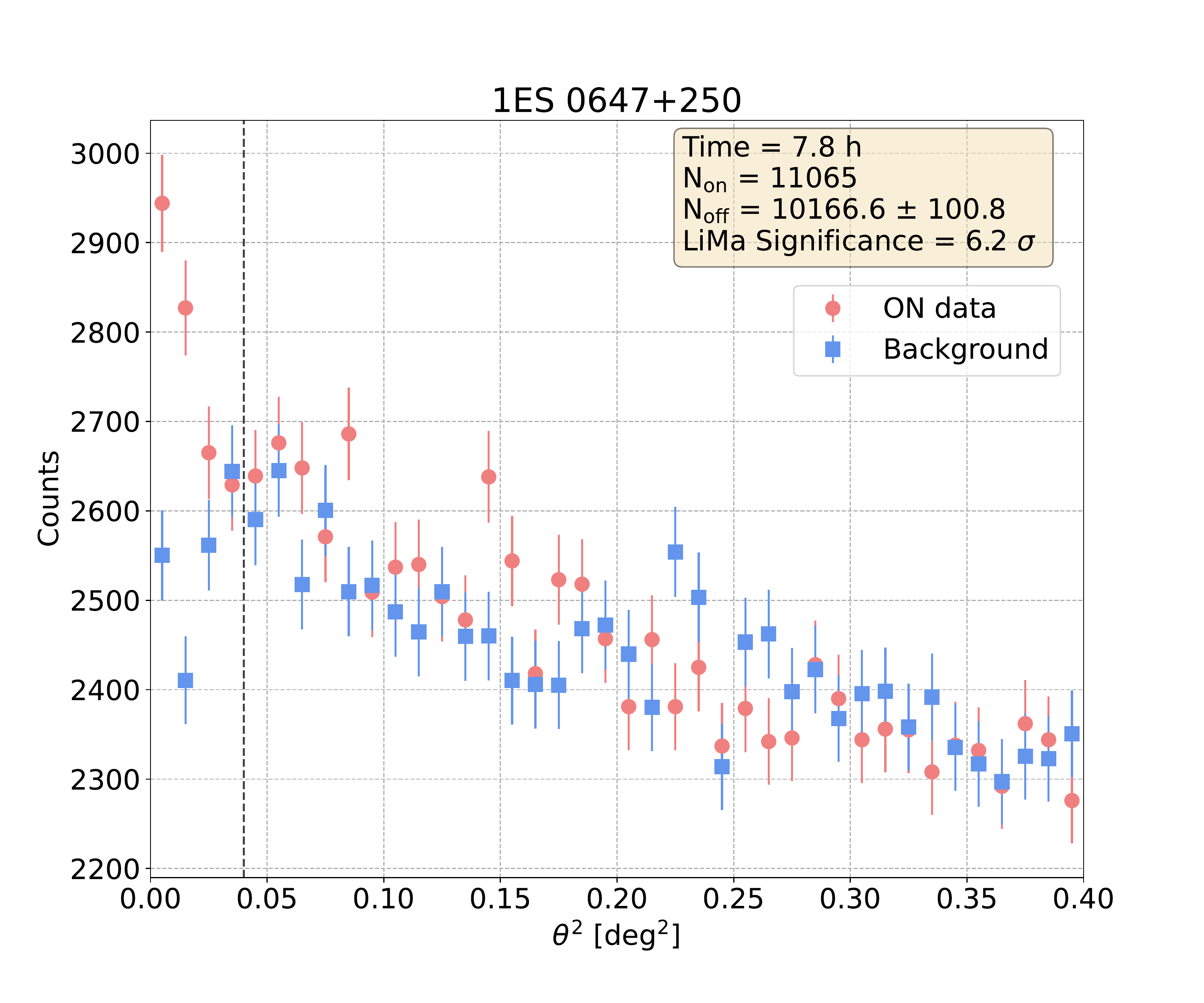}
\includegraphics[width=0.49\columnwidth]{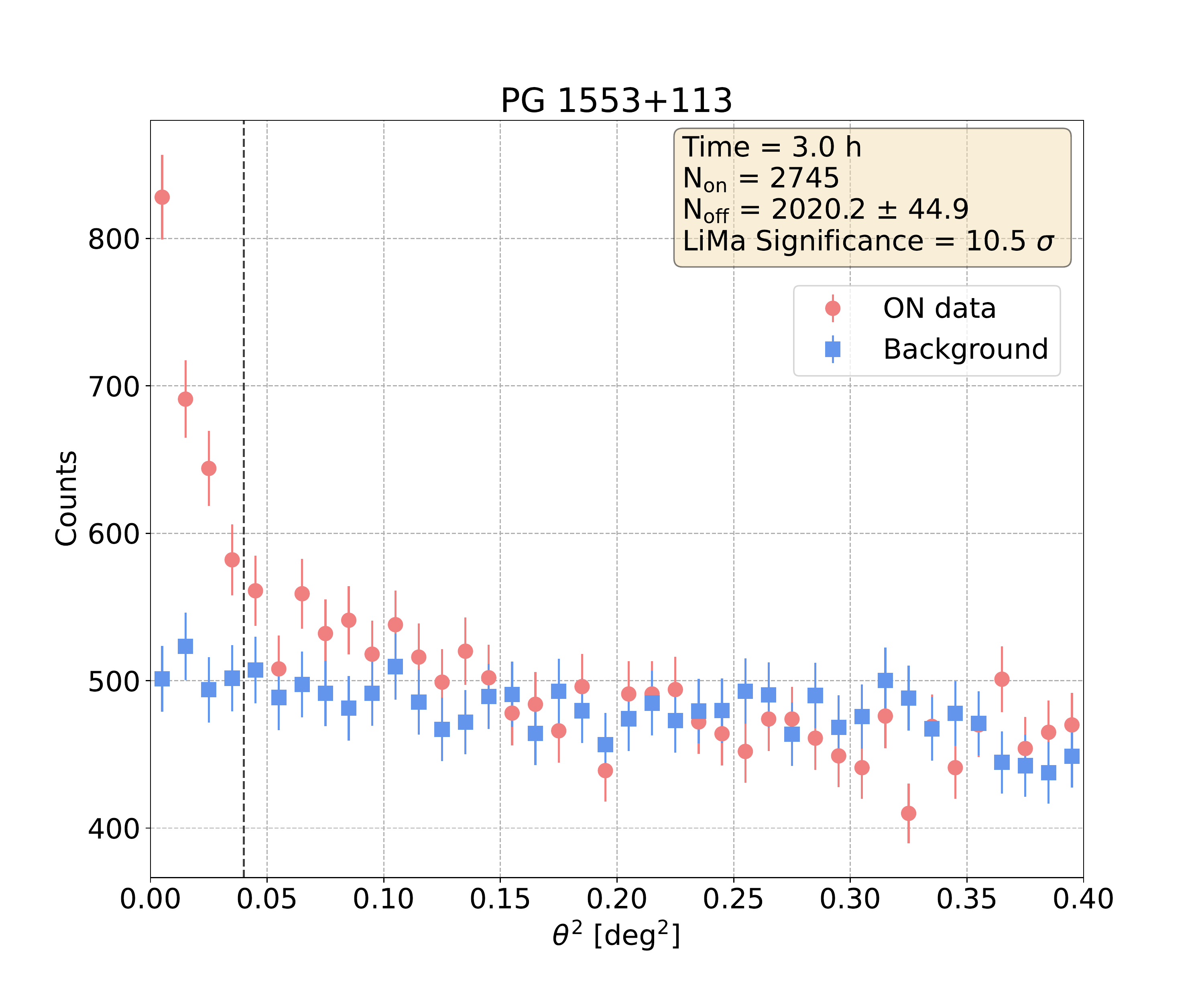}
\caption{$\theta^2$ plots of AGNs detected with LST-1. The $\theta^2$ cut defined is 0.04 deg$^2$.}
\label{fig:agns}
\end{center}
\vspace{-0.7cm}
\end{figure}

AGNs are small regions at the center of a galaxy that emit radiation ranging from radio up to VHE gamma rays. The VHE gamma-ray flux is attenuated via pair production on the extragalactic background light (EBL) photons in the optical and IR bands, which decrease its detectability at high energies with increasing distance. The list of AGNs already detected with LST-1 is the following: Mrk 501, Mrk 421, 1ES 1959+650, 1ES 0647+250 and PG 1553+113. The distribution of the squared angular distance ($\theta^2$) of gamma candidates of the more distant AGNs (1ES 0647+250 and PG 1553+113) is shown in Fig. \ref{fig:agns}.

\subsection{Crab pulsar}
The Crab pulsar (PSR J0534+220) is a young neutron star with a rotational period of 33 ms created after the supernova explosion SN1054. It has a spin-down luminosity of 4.6 $\times$ 10$^{38}$ erg s$^{-1}$ and was first detected at VHE gamma rays by MAGIC \cite{MAGIC_Crab_25GeV} and over the years its spectrum was extended up to TeV energies \cite{MAGIC_Crab_TeV}. We observed the Crab pulsar between January 2020 and March 2021. The observation mode was ON/OFF during the January-February 2020 campaign (for a total of 13.6 hours) and Wobble for the November 2020 - March 2021 campaign (for a total of 30.1 hours). Only data taken under good weather conditions, moonless nights and zenith angle below 35 deg were selected. We applied the source-dependent analysis, which is more sensitive at the lowest energies. The Crab pulsar phases definition is taken from \cite{magic_crab_2012}. The P2 pulse is detected in both data samples, while P1 is marginally significant in the ON/OFF data sample and more significant than P2 in the Wobble one, indicating a lowering of the energy threshold coming from improvements in the trigger settings and the data analysis.


\begin{figure}
\begin{center}
\includegraphics[width=0.8\columnwidth]{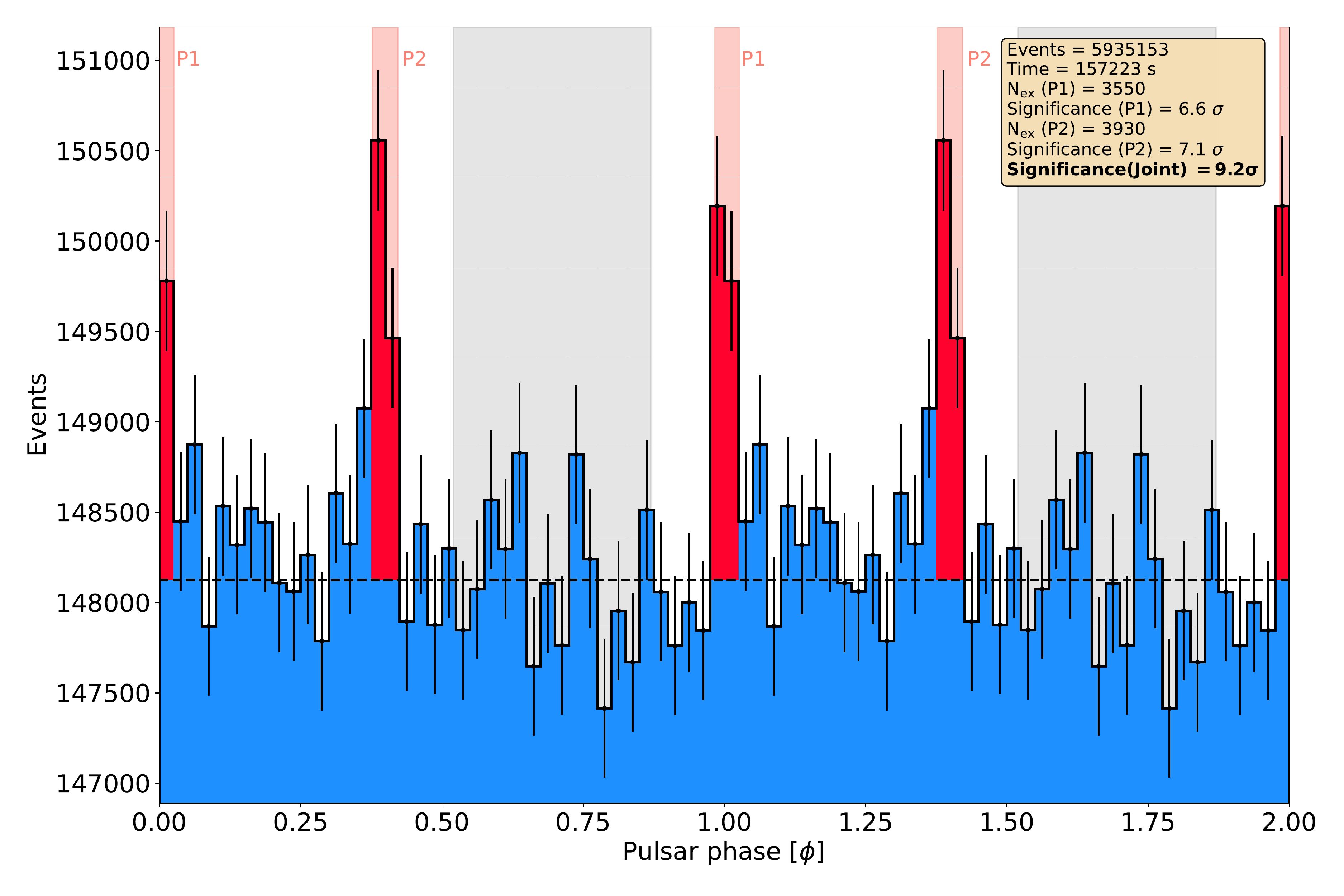}
\vspace{-0.5cm}
\caption{Crab pulsar phaseogram using 43.7 hours of ON/OFF and Wobble observations.}
\label{fig:pulsar}
\end{center}
\end{figure}

\section{Conclusion}

In this proceeding we have shown the physics performance derived using the Crab Nebula observation performed with the LST-1 working in single-telescope mode. We showed the Data/MC comparisons in order to validate the results obtained. We also showed figures of merit like the angular and energy resolution of the telescope, differential sensitivity and Crab Nebula spectra using source-independent and source-dependent analysis. The low energy threshold of the telescope is demonstrated by the observation of several distant AGNs and the detection of significant emission from the two peaks of the Crab pulsar. The detection of the Crab Pulsar with a P1 significance of the order of that of P2 points to an energy threshold in the tens of GeV energy range. 

\section*{Acknowledgements}
We gratefully acknowledge financial support from the agencies and organizations listed here: http://www.cta- observatory.org/consortium\_acknowledgments.

%
%
%

\section*{Full Authors List: CTA LST Project}
\scriptsize
\noindent

H. Abe$^{1}$,
A. Aguasca$^{2}$,
I. Agudo$^{3}$,
L. A. Antonelli$^{4}$,
C. Aramo$^{5}$,
T.  Armstrong$^{6}$,
M.  Artero$^{7}$,
K. Asano$^{1}$,
H. Ashkar$^{8}$,
P. Aubert$^{9}$
% ...
\\

\noindent
$^{1}$Institute for Cosmic Ray Research, University of Tokyo.
$^{2}$Departament de Física Quàntica i Astrofísica, Institut de Ciències del Cosmos, Universitat de Barcelona, IEEC-UB.
$^{3}$Instituto de Astrofísica de Andalucía-CSIC.
$^{4}$INAF - Osservatorio Astronomico di Roma.
$^{5}$INFN Sezione di Napoli.
$^{6}$Aix Marseille Univ, CNRS/IN2P3, CPPM.
$^{7}$Institut de Fisica d'Altes Energies (IFAE), The Barcelona Institute of Science and Technology.
$^{8}$IRFU, CEA, Université Paris-Saclay.
$^{9}$LAPP, Univ. Grenoble Alpes, Univ. Savoie Mont Blanc, CNRS-IN2P3, Annecy.
% ...

\end{document}